\documentclass{elsart}
\newcommand{\brakket}[1]{\langle #1 \rangle}
\newcommand{\tentothe}[1]{ \times 10^{#1}}

\renewcommand{\vec}[1]{\textbf{\textit{#1}}}

\begin{document}             


\begin{frontmatter}
\title{Sonoluminescence: Coupling to an Applied 
Magnetic Field}
\author{B. A. DiDonna, T. A. Witten, and J. B. Young}
\address{Department of Physics and James Franck Institute,
 University of Chicago, Chicago,Illinois 60637}





\begin{abstract}
We investigate several means of coupling between a sonoluminescing 
bubble and an applied magnetic field.  Recent experiments show a 
strong quadratic dependence between the forcing pressures required for 
stable sonoluminescence and magnetic field amplitude. 
However, all coupling mechanisms calculated here for
comparable magnetic fields
involve energies no more than one percent the mechanical
energy of bubble collapse. 
We conclude that the applied 
field must influence the system though its effect on
some parameter which the bubble motion
depends upon very sensitively.
A few such mechanisms are suggested.
{
\setlength{\parindent}{0.0 in}
%

}
\end{abstract}


\end{frontmatter}

\section{Introduction}
\label{sec:intro}

Recent experiments~\cite{young}
show that magnetic fields affect sonoluminescence.
Sonoluminescence is the emission of light during the collapse of 
a bubble in a liquid subjected to an oscillating pressure 
field.~\cite{lit1,lit2,lit3,lit4,lit5}
Young et al.~\cite{young}  found that 
the presence of a uniform magnetic field 
increased the pressure amplitude for both the onset of stable 
sonoluminescence and the ultimate destruction of the bubble at high 
driving forces.  Both thresholds shifted approximately quadratically 
in magnetic field strength, scaling by 1 atm per ${(20 \textrm{T})^2} $.  
This 
compares with typical driving forces of around 1 atm in the absence of 
a magnetic field.

Evidently fields of the order of 10 Tesla are enough to cause substantial 
changes in the emitted light.  This implies a significant coupling of 
the magnetic field energy with the other energies in the collapsing, 
light-emitting bubble---kinetic energy and compressional energy.  The 
detailed calculations of zero-field bubble collapse by Wu and Roberts 
\cite{simulation} provide the quantitative basis for our estimates.  
From these calculations and previous 
experiments~\cite{lit1,lit2,lit3,lit4,lit5} a 
commonly-accepted scenario of the luminescence emerges.  Under optimal 
conditions a bubble is trapped at the antinode of a sinusoidal 
acoustic field near the center of a driven resonant acoustic chamber.  
During a large part of the acoustic cycle, the bubble is unstable 
against expansion.  It expands from a radius of roughly 1 micron to a 
radius of thirty microns or more.  Ultimately the ambient pressure 
becomes sufficiently positive that this large bubble collapses: the 
surface accelerates inward under the unbalanced external pressure.  
This inward motion continues past the point of force equilibrium; the 
inertia of the external liquid continues to compress the bubble.  The 
collapse continues until the interior gas reaches extreme conditions.  
It becomes hot enough to ionize.  Later, the inward velocity exceeds 
the thermal velocities inside the bubble and an imploding shock wave 
is launched when bubble radius is a fraction of a micron.  The emitted 
light is thought to come from the highly compressed center of this 
shock-wave region, over a timescale of picoseconds or less.  
Ultimately the inward motion is stopped by the large excess of outward 
force and then the bubble re-expands.  This re-expansion occurs faster 
than the acoustic period; thus the ambient pressure during expansion 
is still strong enough to cause collapse.  This results in a sequence 
of small collapse bounces after the initial one.  Optimal 
sonoluminescence requires specific choices of gas dissolved in the 
water.  Bubble trapping can be observed without collapse; likewise 
trapping and collapse can be observed without substantial light 
emission.

In this paper we survey the potential coupling mechanisms between the 
magnetic field energy and other forms of energy.  We begin with the 
sonoluminescent plasma, the site of strongest interaction.  After 
estimating the strength of the coupling here, we consider regions 
progressively more distant from the center.  Some have more than one 
possible coupling mechanism.  Our estimates do not claim to cover new 
theoretical ground.  The interaction of magnetic fields with imploding 
matter is a well explored subject in astrophysical
\cite{alfven} and nuclear physics \cite{nuclear} contexts.
Our aim here is to evaluate these well-known effects under the 
specific experimental conditions where sonoluminescence is observed.  
The main coupling of the external field to the ionized interior of the 
emitting bubble arises through its conductivity.  Plausible estimates 
of the conductivity and conducting volume allow one to judge how much 
the plasma modifies the field.  Beyond this conducting volume is a 
neutral region within the bubble but outside the shock front.  We 
consider possible conductivity-based coupling in this region.  
Proceeding outward to the bubble surface, we consider the effects of 
surface charge and of diamagnetic contrast.  Beyond the surface, in 
the bulk liquid, we consider the possibility of Seebeck currents.

All calculations given here are based on the simulated bubble collapse 
detailed in Wu and Roberts~\cite{simulation}.  Although Wu and Roberts 
base their calculations on a bubble of nitrogen gas, we assume an 
argon gas bubble, since argon is very effective at luminescing.  The 
choice of gas is for concreteness, and does not significantly affect 
numerical results.

The magnetic field interaction energies must be compared to the 
zero-field energy of the collapse.  
For all these potential effects the simplest order-of-magnitude
estimates are sufficient to judge their importance.
The hydrodynamic bubble collapse 
powers itself, so that the bubble has absorbed all the mechanical 
energy it will 
receive from the acoustic field at the point when it reaches its 
maximum radius, just before it implodes.  The energy of the collapse 
is then (assuming that the bubble expands adiabatically, i.e.
$ pV^{\gamma} $ is constant, for $\gamma = 1.4$) \( U =\int P 
dV \).  For the bubble cycle simulated in 
\cite{simulation}, the maximum bubble radius is $ 30 \mu$ 
m, so the energy of collapse is 0.03 ergs, or 3 nJ.
 
An independent method of calculating the collapse energy involves 
the surface speed of the bubble in the final moments of implosion, when
all the collapse energy is in the kinetic energy of the inward
rushing water.  Assuming the water is incompressible and therefore the 
velocity of
the fluid scales as $1/r^2$, the kinetic energy in the water is 
$ \int_R^\infty \frac{1}{2} 
\rho_w v^{2}(r)\ dV = 2 \pi \rho_w R_s^3 v^{2}_s $, where 
$R_s$ and $v_s$ are the surface radius and speed, respectively.
At the point when the bubble reaches a radius of one
micron, the surface speed of the bubble wall approaches $10^5$ cm/s,
so the kinetic energy of the water is about $0.06$ ergs.  Our two methods
of calculating the collapse energy agree to a factor of two, which gives 
us a good level of confidence that this is the energy scale typical
of the system.

In all cases the 
mechanisms we investigated appear to be one percent or smaller
perturbations to the mechanical energy of bubble collapse.
We therefore conclude that the applied magnetic field must
be changing some sensitive parameter in the hydrodynamic 
equations which we cannot analyze with our simple energy
scale comparisons.  A possible effect of this nature is 
discussed at the end of this paper.

\section{Interior Plasma}

\subsection{Flux Trapping}
\label{sec:flux}

We first considered the interaction of the external field with the hot 
plasma formed just before the light burst.  The coupling of magnetic 
fields to shock waves in a collapsing bubble has been studied in 
greater detail elsewhere~\cite{chou} for both the high and low limit 
of magnetic Reynolds number.  We content ourselves here to make only
energy estimates based on our ``typical'' sonoluminescence 
parameters. 
 
The collapsing 
plasma compresses the magnetic flux and the collapse is thereby 
inhibited.  The final collapse and formation of a shock wave take 
place over a timescale of $10^{-10}$ seconds and a length scale of 0.3 
microns.  Temperatures can range from $10^3$ to $10^7$ K.  For flux 
trapping to be an important effect, two conditions must be met:
the flux must remain 
trapped for a time comparable to the collapse time and the 
work required to compress the field lines must be comparable to the 
energy in the bubble without magnetic field.

First we calculate the magnetic diffusion time and compare it to the 
collapse time. 
Flux trapping requires that the diffusion time be longer than the
collapse time.
Jackson~\cite{jackson} gives the time constant of magnetic 
field diffusion as
\begin{equation} 
\tau_D = \frac{ 4 \pi \sigma L^2}{c^2} 
\end{equation}
where $\sigma$ is the conductivity and L is the characteristic length 
scale.  We model the conductivity of the plasma in Drude style, giving 
a DC conductivity \( \sigma = {n e^2 \tau}/{m_e} \) where $\tau$ is 
now the mean-free scattering time for an electron in the material and 
$n$ is the density of conduction electrons.

As a first approximation which should overestimate the time constant
and therefore the conductivity, we treat the free electrons as a
Maxwell-Boltzman gas.  The time constant is then given by the mean
free path divided by the average velocity, \( \tau =
{l}/{\brakket{v}} \), \(\brakket{v} \approx \sqrt{{3 k_B
T}/{m_e}} \).  The mean free path is \( l \approx {1}/(\sigma_o N)
\) where $\sigma_o$ is the atomic cross section (Coulombic) $\approx
16 \times 10^{-16}$ cm$^2$.

Labeling \( n = {\mathcal Z} (T,L) N \), with ${\mathcal Z}$ the average 
ionization of the atoms and $N$ the number density of atoms, we find
\begin{equation} 
\sigma = \frac{{\mathcal Z} e^2}{\sigma_o \sqrt{3 m_e k_B T}}
\end{equation}
Therefore the characteristic timescale is:
\begin{equation} 
\tau_D = \frac{4 \pi {\mathcal Z} e^2}{\sigma_o  c^2 \sqrt{3 m_e k_B T}}
L^2 \ \textrm{or} \ \tilde{\tau}_D = 3 \times 10^{-3} \frac{{\mathcal Z}}
{\sqrt{\tilde{T}}} \tilde{L}^2 
\end{equation} 

Here $\tilde{\tau}_D$ is the time constant in seconds,  $\tilde{L}$ 
is the size of the region in cm, and $\tilde{T}$ is the temperature
in Kelvin.

If we take \( \tilde{L} = 3 \times 10^{-5}\) cm, then 
\( \tilde{\tau}_D \approx 3 \times 
10^{-12} {\mathcal Z}/\sqrt{\tilde{T}} \).  
Considering a bubble of argon gas, 
even a massive electrical breakdown at 1000 K would only produce at 
most 8 electrons per atom and therefore a time constant of one 
picosecond.  Better modeling of the ionization using Saha's 
equation~\cite{saha} gives values of \( {\mathcal Z}/\sqrt{\tilde{T}} \) on 
the order of 0.02 for temperatures between $10^3$ and $2 \times 10^5$ 
K, which gives values of $\sigma$ on the order of $10^{16} $s$^{-1}$ and 
timescales on the order of $10^{-13}$ seconds, about 3 orders below 
the timescale of the sonoluminescing shock wave.  The corresponding 
magnetic Reynolds number is $R_{M} = v \tau / L = 0.25$, safely in 
the regime of diffusion dominated (untrapped) field motion.

Next we compare the magnetic field energy to the energy in the 
bubble.  The field energy density for a 10 T magnetic field is $4 
\tentothe{8} $ ergs/cm$^{3}$, so if the plasma couples to the 
magnetic field when the bubble radius is 0.3 microns the bubble 
encloses a field energy of only $ 5 \tentothe{-5}$ ergs.  Thus even 
if there is perfect coupling to the magnetic field, the additional 
field pressure is at most a small perturbation to the 
system.\footnote{It was suggested to the authors~\cite{laughlin} 
that the change in the heat conductance of the
plasma due to the addition of the magnetic field may be enough to
inhibit the shock. To find the effect of the B field
on the heat conduction by thermal electrons,
we must compare the cyclotron radius of the electrons in 
the 10 T field to their mean free path inside the plasma.
The Maxwell-Boltzman value for electron 
velocities at the plasma temperatures given above will approach
$ 10^8 $ cm/s. Classically, the cyclotron radius for an electron
traveling at this velocity is $ r = \frac{m_e vc}{eB} \approx
5 \times 10^{-5} $ cm. 
With atomic number densities of up to $10^{22}$
cm$^{-3}$ in the plasma core, the mean free path for electrons
is only on the order of $10^{-8}$ cm.
So it seems that the heat conductivity contributed by electrons 
in the the plasma is not significantly affected by the 
application of the magnetic field.}

\subsection{Lorentz Forces}

So it seems that the magnetic field is essentially unscreened by the 
ionized plasma.  Accordingly, we consider the forces on the plasma 
caused by the unscreened field.  Lorentz forces will result in a 
current transverse to the collapse velocity. In the frame of the 
collapsing plasma the particles experience an electric field $\vec{E}' 
= \mathbf{\beta} \times \vec{B}$ in Gaussian units, 
where $\beta 
\equiv v/c$ ($= 10^{-4}$ in the plasma collapse).  This results in
a current density (transverse to the collapse velocity) of magnitude 
$\vec{J} = \sigma \vec{E}'$.  The current will drain energy from the 
system in two ways --- it will dissipate energy through ohmic 
heating and store energy through its interaction with the external 
magnetic field.  We calculate the energy of both interactions.

Ohmic heating will dissipate energy at a rate 
\begin{equation}
\label{eq:ohmic}
P = \int_{V} \vec{J} \cdot \vec{E} \approx \sigma \beta^{2} B^{2} V.
\end{equation}
For a field of 10 T 
and a plasma bubble of radius 0.3 microns, the power dissipated is $P 
= 10^{5} $ ergs/sec --- if power were dissipated at this rate for the 
entire duration of the plasma ($10^{-10}$ s), the total energy lost in 
the plasma interaction would be less than $10^{-5}$ ergs.  Thus
Ohmic loss due to the plasma-magnetic field interaction 
has no significant effect on the bubble motion.

To find the interaction energy between the current and the applied 
magnetic field we calculate the magnetic moment of this current 
distribution, taking it to be mainly dipole by axial symmetry.  
Defining the z axis in the direction of the magnetic field, The net 
magnetic moment is
\begin{eqnarray}
\label{eq:dipole}
\vec{M} & \equiv & \frac{1}{2c} \int \vec{x}' \times \vec{J}\ d^{3}x' 
 \\
& = & \frac{\pi}{3} \frac{ \sigma v B}{c^{2}} R_{p}^{4} \ \hat{z}, 
\nonumber
\end{eqnarray}
where we have assumed that $v$ is essentially constant throughout the 
plasma. Here $R_{p}$ is the radius of the plasma bubble, 0.3 microns.
The first order correction 
to the field energy $U = \frac{1}{2} (\vec{B} + \vec{M})^2 $ 
due to the induced magnetic moment is then
\begin{equation} 
U = \vec{M} \cdot \vec{B} = \frac{\pi}{3} \left( \frac{ \sigma 
v}{c^{2}} \right)  R_{p}^{4}  B^{2} 
\end{equation}
For the plasma parameters given previously, the interaction energy 
amounts to $3 \times 10^{-7} $ ergs, which is much smaller than 
other energy scales in the sonoluminescence cycle.
Thus it is clear that the hot plasma at the core of the sonoluminescing 
bubble does not couple to external magnetic field strongly enough to 
alter the local field or the gas dynamics significantly.  We therefore 
move on to examine coupling in other regions of the fluid-bubble system.

\section{Non-Plasma Bubble Interior}

Another possible locus of the magnetic field effect is the region 
outside the plasma.  Immediately outside the plasma is an un-ionized 
region of the bubble.  Due to the scarcity of free charge in the 
region, we saw no likely prospect for a strong magnetic field effect 
there.  The one possible source of current in this region is the large 
thermal gradient between the plasma core and the bulk fluid outside 
the bubble.  However, this subject is treated later for currents 
outside the bubble,  where it is shown that the thermal currents 
required for significant coupling are greater than any we would expect 
to find in a system this small.  We therefore don't consider 
any magnetic field interactions in the non-plasma bubble interior.

\section{Bubble Surface}

\subsection{Surface Currents}

There is a possibility of finding large currents confined to the 
surface of the bubble, since there will be a localized ion build-up at 
the gas-fluid interface. This greatly enhances the electrical 
conductivity in that region.  As the surface of the collapsing 
bubble sweeps through the applied magnetic field, Lorentz forces will 
cause a current to flow around the bubble.  We compute 
the energy associated with this current and compare it to the 0.03 
ergs which the bubble absorbs from the sound field (c.f.  Section 
(\ref{sec:intro})).  In order to estimate the maximum possible 
coupling to the field, we estimate the maximum carrier density and 
conductivity of the surface.

To estimate the maximum effect of the magnetic field we treat the 
bubble surface as a
room temperature plasma.  A typical charge density for a fully ionized 
surface is $0.2 $ C/m$^{2}$, or $ 1.25 \tentothe{14} $ 
(charge carriers)/cm$^{2} $ ~\cite{surface}.  It would be 
generous to assume that the carrier density at our bubble surface 
corresponds to approximately twice the surface charge given above 
(ions and their counterions) contained in a thin shell surrounding the 
bubble surface.  
Because of the small volume 
and low temperature of the charged shell, we do not expect strong 
coupling to the magnetic field (flux trapping), so we confine these 
calculations to the limit of unscreened magnetic field.  The 
mechanisms for energy loss are exactly the same as those calculated 
above for the interior plasma bubble --- Ohmic loss and field 
interaction energy.  In either case we must estimate the electrical 
conductivity at the surface.

Since the current is made up of ions (dissociated water or salts) 
migrating through the liquid, the ion mobility is controlled by 
viscous drag.  The maximum mobility for a sphere in a viscous medium 
is \( v/F = {1}/{(6 \pi \eta a)} \), where in this case $\eta$ is the 
viscosity of water ($0.01$ g cm$^{-1}$ s$^{-1}$) and $a$ is the 
effective radius of the ions, which we take to be about 2 \AA.  
Substituting this into the definition of conductivity, $ \vec{j} = 
\sigma \vec{E} $, and assuming most carriers are singly ionized, 
so that $ \vec{j} = n e \vec{v} $ and $ \vec{F}_{ion} = e \vec{E} $, 
we find
\begin{equation}
\sigma = \frac{ne^{2}}{6 \pi \eta a}.
\end{equation}
If we assume the current flowing over the surface of 
the bubble is confined to a shell of thickness $\textit{w} \approx 10$ 
\AA or less, then the 
local charge density $n$ is approximately $2.5 
\tentothe{18} $ (charge carriers)/cm$^{3}$ and
the conductivity is $10^{8}$ s$^{-1} $ (for comparison, typical metallic 
conductivities are on the order of $10^{18}$ s$^{-1}$ and the 
conductivity of pure water is $4.5
\tentothe{4}$ s$^{-1}$).

Using Eq.(\ref{eq:ohmic}) to calculate the Ohmic dissipation, we find 
the power loss is $P = 4 \pi \sigma (\dot{R}/c)^{2} B^{2} R^{2} \, 
\textit{w}$.  (Note that the $\sigma$ calculated above is proportional 
to $1/\textit{w}$ through its dependence on $n$, so the combination 
$\sigma \, \textit{w}$ is independent of $\textit{w}$ and the 
equations given here are valid for any $\textit{w} \ll R$.)  The 
maximum surface speeds are on the order of $10^{5}$ cm/s, the time 
averaged bubble radius over the period of the collapse will be less 
than 20 microns, and the collapse itself takes about $2 \ \mu$s.  So a 
bubble collapse in the presence of a 10 T external field would 
dissipate less than $10^{-11}$ ergs --- nine orders of magnitude less 
than the mechanical energy in the bubble.

Next we calculate the energy in the dipole interaction between current 
and magnetic field.  Starting from Eq.(\ref{eq:dipole}) we take the 
dipole moment of the thin conducting surface to be \( M = \frac{4\pi}{3}
\frac{ \sigma \dot{R} B}{c^{2}} R^{3} \textit{w}
\ \hat{z}. \) The field interaction energy is then \( 
U = (\frac{ \sigma \textit{w} \dot{R}}{c^{2}}) \frac{4 \pi}{3} R^{3} 
B^{2} \). The factor in parentheses is dimensionless and independent 
of $\textit{w}$, the factor after 
it is proportional to the magnetic field energy enclosed by the 
bubble. In a 10 T field, the bubble encloses a field energy of about 1000 
ergs 
at maximum radius (30 microns); however, the surface velocity at this 
point is about $10^{3}$ cm/s, so the term in parentheses is $ 10^{-17} 
$ and the total energy only $10^{-14}$ ergs.  Later in the collapse
surface speeds are on the order of of $10^{5}$ cm/s, but by 
this time the radius is 10 microns or less, so the total energy 
from above is at most $ 4 \times 10^{-17}$ ergs.  
This is nowhere near the 0.03 erg total energy of the collapsing bubble.  
From 
these numbers it is clear that the energy of the plamsa-field interaction
at the bubble surface is no more than $10^{-9}$ of the nonmagnetic 
bubble energy.

\subsection{Diamagnetic Pressure}

A further source of magnetic effect is not through Lorentz forces but 
through the magnetic polarizability $\chi$ of the fluid.  The 
diamagnetism of water must add an additional outward pressure to the 
bubble walls.  The magnetic permeability is defined as
$\mu = 1 + 4\pi \chi$.
According to~\cite{magfluid}, the difference in $\chi$ across the 
surface adds a uniform outward pressure on the bubble wall of
order $\Delta \chi B^2$.
For pure water, $\chi = -7.2 \times 10^{-7}$ so the pressure 
on a bubble in water at $10$ Tesla will be $\sim 3.6 
\times 10^3$ dynes/cm$^2$ ($3.6 \times 10^{-3}$ atm).  
The field will also create a non-uniform pressure at the bubble
surface that goes as $\left( \Delta \chi \right)^2 
\left|\vec{B} \cdot \vec{n} \right|^2$, which in our case is 
seven orders smaller than the uniform component.
(We neglect 
the susceptibility of the gas inside the bubble. When the bubble is
at maximum radius, the gas inside is so dilute that the diamagnetic
susceptibility of water would be greater by a factor of two than 
even the paramagnetic susceptibility of a pure O$_{2}$ bubble. Since
the results of Young et. al. were identical for pure argon bubbles, 
the susceptibility of the gas cannot be making any significant
difference.) 
We note that this 
is a lower bound, since typical magnetic impurities in the water could 
increase its susceptibility and enhance this effect.  The magnetic 
impurity concentration of the water used in Kang's experiment is not 
known.  

The diamagnetic pressure given above should be compared with the 
pressure on the bubble wall at maximum radius, which, as noted in the 
introduction, is a good measure of the total energy deposited in the 
bubble by the acoustic field.  In the simulation by Wu and Roberts, 
the pressure at maximum radius is 0.54 atm, so the diamagnetic 
pressure perturbs the mechanical energy by one part in a hundred. 

This small perturbation may be significant, since
even a slight change in water pressure can cause
a marked change in the concentration 
of dissolved gases in the bubble.
It was shown in~\cite{ambpressure} that a 5~\% change
in the ambient pressure of the surrounding water can decrease the 
SL light intensity by 200~\%, due to the change in saturation 
concentration of argon gas in the fluid.
The $\sim 1$~\% pressure drop due to diamagnetism could well have
similar large effects on the gas concentration.
The gas solubility effect cannot be analyzed within
the energy scale framework set out in this paper, so we leave 
the possibility as a subject of future investigation.

Fortunately, the susceptibility effect can be 
tested by adding a paramagnetic salt to the water.  A 0.05 molar 
solution of manganese chlorate should balance the outward diamagnetic 
pressure with an inward paramagnetic pressure.  A 5 molar solution 
should produce a magnetic field pressure 100 times stronger than that 
in pure water.  The nature of the paramagnetic effect may be 
different, since the direction of the pressure drop is reversed.  
Also, the presence of dissolved salts will alter
the gas solubilities even in the absence of field.
This test remains to be done.

\section{Bulk Fluid}

\subsection{Thermally Induced Currents}
\label{sec:seebeck}

Though we expect no significant currents on the bubble surface, there 
may be more extended currents in the water outside the bubble.  This 
region has only weak coupling to the field (relative to the plasma and 
the bubble surface).  On the other hand, it has a large volume.  How 
could the field couple to the matter in this region?  If it is to 
couple via the Lorentz force, there must be electric currents.  The 
most obvious source of such currents we anticipate is a thermal 
gradient.  The occurrence of such currents is known as the Seebeck 
effect. Accordingly, we asked whether a thermally generated current in 
the surrounding water, interacting with the $B$ field via the Lorentz 
force, might convert a significant amount of the inward momentum of 
the collapsing bubble into angular momentum.

The Seebeck coefficient of our water would enable us to convert this 
temperature drop to an electric current.  We do not know this 
coefficient for the water used in our studies.  However, we can attack 
the problem from the other end.  We ask what sort of outward electric 
current $I$ would be required to substantially alter the momentum of 
the water?  If the current required for strong magnetic coupling is 
much greater than any current that could exist in the system, then the 
coupling between the actual Seebeck currents and the external magnetic 
field must be negligible.  To estimate the order of magnitude of the 
current required to strongly affect the bubble collapse, we demand 
that the interaction of this current with the external magnetic field 
be enough to add a transverse momentum equal to all the inward 
momentum of the collapse.  Remembering that a significant
Seebeck current will only flow for a small period of  time 
$ \Delta t $ when
there is a large temperature gradient, we set
\begin{equation}
\label{eq:sbeq}
\int_{t_{o}}^{t_o+\Delta t} dt'\int_{r>R} d^3 r 
\vec{j}(r, t') \times \vec{B} 
  \approx \rho_{water} \int_{r>R} d^3 r v(r, t_{o})
\end{equation}

  Again we use liberal 
estimates so as to find the maximum possible effect.
The greatest current should flow at the last moment of the 
hydrodynamic collapse, a period $\Delta t$ lasting about $1$ ns, 
when the 
temperature of the bubble core jumps by a few orders of magnitude.  We 
consider for definiteness the current flowing in the region of fluid 
between the bubble surface and twice that radius.  At this time in the 
collapse the typical bubble radius is around 1 micron, so we consider 
the current flowing radially outwards across a distance $\Delta r = 1 
\ \mu$m, filling a volume \( \Delta V \approx
\frac{4\pi}{3} (2 \times 10^{-6} \textrm{m})^3 \). 
Typical bubble surface speeds 
are on the order \(v = 10^3 \) m/s. 
We can therefore approximate the equality in Eq.(\ref{eq:sbeq}) by:
\begin{equation} 
\rho_{w} \Delta V v \approx IB\Delta r \Delta t. 
\end{equation} 
With \(\rho_{w} = 10^3 \) kg/m$^3$ and for a 10 T applied field, solving 
for the current gives \(I = 3 \times 10^3 \)amps. Let us put a 
current of this magnitude in perspective by calculating the associated 
ohmic heating. Very pure water has a 
resistivity of \(\rho \approx 2 \times 10^5 \) $\Omega$-m; relatively 
impure water solutions can have resistivities as low as 1 $\Omega$-m.  
Using the lower resistivity, a current 
of this magnitude flowing across a distance of $\Delta r = 1$ micron 
would encounter a total resistance of $R = \rho \Delta r / 
\mbox{\emph{(area of bubble)}} \geq 2 \times 10^{4} \ \Omega$ and 
therefore dissipate \( I^{2} R \ \Delta t \geq 2 \times 10^{9} \) 
ergs as heat.  This is clearly more energy than is deposited in the 
bubble, and much more than we would expect to find localized in any 
area in the system.  We conclude that magnetic perturbation via 
Seebeck currents must be negligible.

\section{Conclusion}

Our survey of possible magnetic effects has not identified a mechanism 
with an associated energy comparable with that of the unperturbed 
bubble. 
It is clear, however, that the magnetic field couples directly to
many of the parameters involved in the hydrodymanics
of bubble motion.  We can only conclude that 
although the field does not directly
change any hydrodyamnic parameter by a factor of order
unity, it must significantly change some parameter upon
which the system depends very sensitively. 
Under this assumption, one possible explanation
of the magnetic field's effect on the system is through
its influence on
the concentration of dissolved gasses inside the bubble. 
The influence of dissolved gas concentrations on SL
has already been studied~\cite{ambpressure}; it would be 
relatively straightforward to perform a detailed calculation
of the partial pressure dependence on applied magnetic field.

We would like to thank M. Brenner, S. Hilgenfeldt, W. Kang, and
R.B. Laughlin for 
enlightening discussions and suggestions.
This work was supported primarily by the MRSEC Program of the National 
Science Foundation under Award Number DMR-9400379

\end{document}